\documentclass[aoas,nameyear,MSNbibl,dvips]{arximspdf}
\usepackage{breakurl}

\doi{10.1214/10-AOAS398K}
\referstodoi{10.1214/10-AOAS398}
\volume{5}
\issue{1}
\pubyear{2011}
\firstpage{80}
\lastpage{82}

\makeatletter
\newcommand{\bld}[1]{\mathbf{#1}}

\makeatother

\begin{document}
\begin{frontmatter}

\title{Discussion of:  A statistical analysis of multiple temperature proxies: Are
reconstructions of surface temperatures over the last 1000~years~reliable?\thanksref{T1}}
\runtitle{Discussion}
\pdftitle{Discussion on A statistical analysis of multiple temperature proxies:
Are reconstructions of surface temperatures over the last 1000 years reliable?
 by B. B. McShane and A. J. Wyner}
\begin{aug}
\author[A]{\fnms{Doug} \snm{Nychka}\corref{}\ead[label=e1]{nychka@ucar.edu}}
\and
\author[B]{\fnms{Bo} \snm{Li}\ead[label=e2]{boli@purdue.edu}}

\runauthor{D. Nychka and B. Li}

\affiliation{National Center for Atmospheric Research and Purdue University}

\address[A]{National Center for Atmospheric Research\\
Institute for Mathematics Applied\\
to Geosciences\\
1850 Table Mesa Drive\\
Boulder, Colorado 80305\\
USA\\
\printead{e1}} 

\address[B]{Department of Statistics\\
Purdue University\\
250 N. University Street\\
West Lafayette, Indiana 47907-2066\\
USA\\
\printead{e2}}
\end{aug}
\thankstext{T1}{Supported by NSF Grant ATM-07-24828 and DMS-10-07686.
The National Center for Atmospheric Research is managed by the
University Corporation for Atmospheric Research under the sponsorship
of the NSF.}

\received{\smonth{9} \syear{2010}}
\revised{\smonth{9} \syear{2010}}



\end{frontmatter}

This article (MW) has stimulated much valuable discussion and helped to
focus attention on an important area for the application of statistics.
Given the short amount of space, however, we reluctantly comment only
on the second and last sections.

\section*{Excursions in the history of science}
Although Section 2 of this paper is lively reading, we feel that the
viewpoint is not balanced and emphasizes statistical correctness over
the broader issues of scientific understanding. Recounting a
controversy that has both a political dimension and involves scientific
issues from several disciplines is perhaps better left to a historian
of science. Wegman's quote on page 9 of the
article is actually from a later written response to Representative
Stupak, not from the original testimony [see Questions surrounding the
hockey stick (\citeyear{Q2006})]. We encourage readers to also read the transcript
of the congressional hearings and the contemporaneous report by the
National Academies, NRC (2006) to follow this debate.

\section*{Paleoclimate reconstructions}
The Wegman committee's original report stopped short of redoing the
temperature reconstruction with Mann's data and with the correct
centering of the principal components. Although this exercise was
beyond the report's charge, it is sound statistical practice to
evaluate changes in intermediate methodology by their influence on the
final statistical inference. The string of references that are cited by
MW on page 10 beginning with Mann and
Rutherford (\citeyear{MR2002}) established the robustness of the reconstruction with
respect to centered verses noncentered methods if several PCs are
included. This is a finding that might have been uncovered by the
Wegman committee as well. In this context, we applaud MW for carrying
through to a reconstruction to assess the impact of methodological
choices. We term the model used in Section 5 a \textit{direct} approach
because it builds a predictive regression model for temperature
directly from the proxies. To complement this article, we discuss an
\textit{indirect} approach that takes advantage of some current work in
Bayesian statistics.

\section*{A Bayesian hierarchical model (BHM)}
Although a direct approach may be useful for comparison with previous
work, we hold that a BHM provides a better solution to the
reconstruction problem. A BHM can be described as \textit{indirect} in
that one models the dependence of the proxies conditional on
temperature. Bayes' theorem is then used to invert the relationship to
arrive at a posterior predictive distribution of temperature given the
data. We sketch this approach below using seminal ideas from Tingley
and Huybers (\citeyear{Tingley2010}) and some features from Li, Nychka and Ammann (\citeyear{Li2010}).
Let $\bld T_t$ be the true temperatures on a grid at time $t$ and let
Northern Hemisphere (NH) temperature, $y_t$, be a linear combination of
the $\bld T_t$.
A possible HBM for this problem is:\vspace*{12pt}

\noindent
{\fontsize{9pt}{11pt}\selectfont{
\tabcolsep=0pt
\begin{tabular*}{\textwidth}{@{\extracolsep{\fill}}lcc@{}}
\hline
\textit{Data level:}& Proxies  & $x_{t,i} = \gamma_i  \mathbf{h}_i \mathbf{T}_t + u_{t,i}$ \\
  \textit{Process level:} &{Space--time process:} & $\bld T_t = y_t \bld1 + \bld v_t $;
$\bld v_t = A\bld v_{t-1} + \bld e_t ; \bld e_t \sim N(0,\Sigma)
$\\
&{NH mean process: } & $y_t = \mu+ S_t \omega_S + V_t \omega_V + C_t
\omega_C + w_t $\\[3pt]
  \textit{Prior level:}
& & $ [ \bolds\gamma, \bolds\omega, A, \Sigma,\ldots   ] $ \\
\hline
\end{tabular*}}}\vspace*{12pt}

The data equation asserts that the $i$th proxy at time $t$ is a linear
combination of the true temperature field plus noise. $\bld h_i$ is a
known row vector of weights and $\gamma_i$ an unknown parameter to
``calibrate'' each proxy. The errors, $u_{t,i}$, for each proxy ($i$)
may have autocorrelation but we will assume that between proxies the
noise time series are independent---the goal is to explain correlation
among proxies by the temperature field (or other geophysical variables).
At the Process level the temperature field evolves as a space--time
process with the variation in the NH average prescribing its mean level.
Here $\bld v_t$ is assumed to be a first-order vector autoregressive
process with
$A$ and $\Sigma$ determining the spatial dependence.
The NH mean level reflects the basic energy balance of the Earth's
climate system. The external series of solar radiance ($S$), volcanic
dust ($V$) and carbon dioxide concentrations ($C$) are large scale drivers
of temperature. A low-order autoregressive process, $w_t$, reflects
additional interannual variation. Finally, the prior level favors
diffuse priors on unknown statistical parameters but can also
consolidate information across many similar parameters. Specifically,
priors for the regression parameters, $\{ \gamma_i \}$ can borrow
strength across proxies and control overfitting. Given this hierarchy,
one samples the predictive distribution for $y_t$ using Bayes' theorem
and Markov chain Monte Carlo.

\section*{Benefits of the hierarchical approach}
It is better to model how a proxy depends on observed climate rather
than formulating a prediction model through a direct relationship.
Climate scientists working on a particular proxy spend much effort in
understanding and
quantifying this forward relationship: conditional on the climate what
would be the response of the proxy? Thus, the data level is a useful
framework to incorporate their expert knowledge into the
reconstruction. The process level is attractive to geoscientists as
well because it builds in constraints in the reconstruction that are
reasonable and well accepted. One strategy for formulating this process
model is to use the output from high resolution climate system models
to identify the form of the vector autoregression ($A$) and the spatial
correlations among the innovations ($\Sigma$).
The BHM addresses missing and irregular proxy information and
temperatures in a consistent way. The posterior distribution can be
sampled when proxies are missing over different time periods and so a
single statistical model is used to derive the reconstruction at all
times. This is in contrast to the direct approach where one has to use
a separate model for each different subset of proxies that are
available for a given reconstruction period.

The hierarchical model and the indirect approach avoid the problem of
proxy centering that was first encountered by Mann. Direct approaches,
even using the Lasso, can suffer from the attenuation effects caused by
measurement errors in proxies [Ammann, Genton and Li (\citeyear{Ammann2010})]. If this
effect is not corrected, the RMSE could be misleading. For example, the
in-sample mean could have smaller RMSE than a biased reconstruction due
to attenuation. However, the biased reconstruction could capture the
basic structure of the temperature process while the in-sample mean
contains no information. In contrast, the hierarchical models with the
indirect approach are free of this concern. Overall, we believe that
HBM are a success in transferring mainstream statistical ideas to a
substantial application in the geosciences and we thank the authors
again for initiating this discussion.


%

\printaddresses

\end{document}